\documentclass[aps,prb,showpacs,groupedaddress,twocolumn]{revtex4}
\usepackage{graphicx}
\usepackage{bm}
\usepackage[bf,tiny,center,indentafter]{titlesec}
\usepackage[CJKbookmarks=true,dvipdfm,colorlinks=false,
pdfstartview=FitH]{hyperref}

\begin{document}

\title{The Effect of the Disorder on the Longitudinal Resistance of a
Graphene p-n Junction in Quantum Hall Regime}
\author{Jiang-chai Chen$^1$, T. C. Au Yeung$^2$, and Qing-feng Sun$^{1,*}$}

\address{
$^1$Beijing National Lab for Condensed Matter Physics and Institute
of Physics, Chinese Academy of Sciences, Beijing 100190, China\\
$^2$School of Electronic and Electrical Engineering, Nanyang
Technological University of Singapore, 639798, Singapore}

\begin{abstract}
The longitudinal resistances of a six-terminal graphene p-n junction under a
perpendicular magnetic field are investigated. Because of the chirality of
the Hall edge states, the longitudinal resistances on top and bottom edges
of the graphene ribbon are not equal. In the presence of suitable disorder,
the top-edge and bottom-edge resistances well show the plateau structures in
the both unipolar and bipolar regimes and the plateau values are determined
by the Landau filling factors only. These plateau structures are in
excellent agreement with the recent experiment. For the unipolar junction,
the resistance plateaus emerge in the absence of impurity and they are
destroyed by strong disorder. But for the bipolar junction, the resistances
are very large without the plateau structures in the clean junction. The
disorder can strongly reduce the resistances and leads the formation of the
resistance plateaus, due to the mixture of the Hall edge states in virtue of
the disorder. In addition, the size effect of the junction on the
resistances is studied and some extra resistance plateaus are found in the
long graphene junction case. This is explained by the fact that only part of
the edge states participate in the full mixing.
\end{abstract}

\maketitle

\address{
$^1$Beijing National Lab for Condensed Matter Physics and Institute
of Physics, Chinese Academy of Sciences, Beijing 100190, China\\
$^2$School of Electronic and Electrical Engineering, Nanyang
Technological University of Singapore, 639798, Singapore}

\titlespacing{\section}{0pt}{10pt}{*4}

\section{Introduction}

The successful fabrication of graphene, a monolayer of carbon atoms
arranged hexagonally, \cite{sci306-666,nat438-197, nat438-201}had
fueled many experimental and theoretical works. For undoped 2-D
graphene sheet, Fermi energy is located at the Dirac neutral point.
Around the Dirac point, the graphene has a linear dispersion
relation, which leads to quasiparticles obeying the massless
Dirac-like equation and presents extraordinary
properties.\cite{rmp81-109,rmp80-1337,prb53-2449} For example, for a
graphene under a strong perpendicular magnetic field, its Hall
plateaus assume half-integer values,\cite{nat438-197, nat438-201,
natmat6-183} $h/[g(n+1/2)e^{2}]$, where $g = 4$ is the spin and
valley degeneracy and $n$ an integer. By varying the gate voltage,
both carrier type and concentration of the graphene sheet can be
tuned.\cite{prl98-236803} A graphene p-n junction is formed by
connecting up one p-type graphene and one n-type graphene. Many
exciting phenomena closely related to the massless Dirac character
of carriers,\cite{natphys2-620, prl102-026807, sci315-1252,
natphy3-172, natphy3-192} such as relativistic Klein tunneling
\cite{natphys2-620, prl102-026807} and Veselago lensing
\cite{sci315-1252}, are predicted for graphene p-n junctions.

Recently, the electron transport through the graphene p-n or p-n-p
junctions was extensively investigated both experimentally and
theoretically.\cite{sci317-638, nanolett9-1973, prl99-166804,
prb79-195327, sci317-641} In quantum Hall regime, Williams \emph{et
al.} \cite{sci317-638} experimentally found that the two-terminal
conductance exhibits plateaus with half-integer values,
$g(|n|+1/2)e^{2}/h$, in the case of unipolar junctions and
fractional values for bipolar junctions. At about the same time, the
theoretical works by Abanin and Levitov \cite{sci317-641} explained
the appearance of the fractional plateaus by means of the mixture of
the electron-like and hole-like Hall edge states in the vicinity of
the junction boundary. There were some subsequent investigations on
graphene junctions. Using Anderson short-range disorder potentials,
Long \emph{et al.} \cite{prl101-166806} and Li \emph{et al.}
\cite{prb78-205308} numerically computed and analysed the transport
behavior of graphene junctions. They found that conductance plateaus
emerge in the case of suitable disorder strength. Also, T. Low
\cite{prb80-205423} considered the long-range interface and edge
disorders in the armchair, zigzag, and antizigzag edge graphene
ribbons, and numerically simulated the result of the conductance
plateaus.

Very recently, Lohmann \emph{et al.} \cite{nanolett9-1973} measured the Hall
and longitudinal resistances in a six-terminal graphene junction device. In
their experiment, the difference of carrier concentrations between two
adjacent regions (the left and right regions) is introduced by chemical
doping, and a global gate voltage controls the carrier concentrations in the
two regions. By tuning the gate voltage and doping densities, the graphene
ribbon can be of p-p, p-n, n-p, or n-n type. They found that Hall
resistances of the left and right regions exhibit half-integer plateaus, as
usual. Furthermore, the longitudinal resistances also exhibit plateau
structures. In particular, the longitudinal resistances at opposite edges
are not equal. For instance, with the filling factors $(\nu _{L},\nu
_{R})=(2,-2)$ (here $\nu _{L/R}$ is the Landau filling factor in the
left/right region), the longitudinal resistance at one edge is $h/e^{2}$ but
it is zero at the other edge. By using the concept of the mixture of Hall
edge states near the p-n junction boundary, they explained the appearance of
these plateaus and the difference between the longitudinal resistances at
the opposite edges. So far, there is not any theoretical investigation which
gives quantitative and numerical result to explain Lohmann's experimental
result. More effort needs to be done in order to find out how disorders
affect the resistance plateaus and how the plateaus depend on disorder
strength.

\begin{figure}[tbp]
\centering
\includegraphics[bb=145 203 445 589,width=8.5cm]{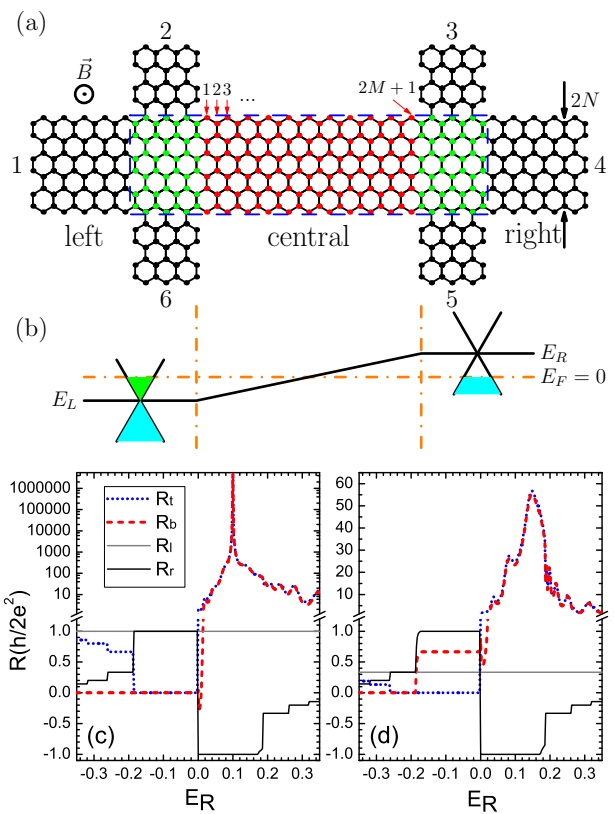}\newline
\caption{ (Color online) The panel \textbf{(a)} is the schematic
diagram for the six-terminal graphene p-n junction device and the
panel \textbf{(b)} is for the Dirac energy $\protect\epsilon_i$ in
the left, central and right regions. The panels (c) and (d) show the
resistances vs. $E_R$ at the clean graphene junction ($W=0$) with
$E_L=-0.1$ \textbf{(c)} and $-0.2$ \textbf{(d) }, respectively. The
other parameters are the width $N=50$, the length $M=20$, and
$\protect\phi=0.01$.} \label{fig1}
\end{figure}

In this paper, we theoretically and numerically study electron
transport through graphene junctions. Following the experiment by
Lohmann \emph{et al.}, we consider the six-terminal graphene
junction device [shown in Fig.~\ref{fig1}(a)]. A perpendicular
magnetic field $B$ is applied to the graphene sheet. By using the
tight-binding Hamiltonian and the Landauer-B\"{u}ttiker formulism in
the framework of non-equilibrium Green's function method, the
longitudinal and Hall resistances are calculated. The numerical
results show that the longitudinal resistances $R_{t}$ and $R_{b}$
of the top and bottom edges, respectively, are usually different in
the presence of the magnetic field, as expected from the property of
the chirality of Hall edge states. There is an essential difference
in the transport behavior between unipolar and bipolar graphene
junctions. For unipolar (n-n and p-p) junctions, the longitudinal
resistances have plateau structures in the case of clean (no
disorder) graphene. The resistance plateaus can keep in the moderate
disorder strength, and they are destroyed until the very strong
disorder. On the other hand, for bipolar (n-p and p-n) graphene
junctions, the longitudinal resistances $R_{t}$ and $R_{b}$ are very
large in the clean case. But in the presence of disorder, they are
strongly decreased even when the disorder strength is weak. For n-p
junctions, the top-edge resistance $ R_{t}$ reduces to a moderate
positive value, but the bottom-edge resistance $ R_{b}$ drops to
zero or even turns negative. In a suitable range of disorder
strength, the resistances $R_{t}$ and $R_{b}$ of bipolar junctions
also have plateau structures due to the full mixing of Hall edge
states. For the lowest filling factors, the resistance plateaus
exist in a very broad range of disorder strength. Hence, they are
produced easily in experiment. But for high filling factors, the
plateau only emerges in a narrow disorder range, if it exists.
Furthermore, for moderate disorder, the plots of the longitudinal
resistances versus the gate voltage exhibit plateau structures in
both cases of unipolar and bipolar junctions and the plateau values
are only determined by the filling factors $(\nu _{L},\nu _{R})$,
which are in excellent agreement with the recent experiment by
Lohmann \emph{et al.}. In addition, we also find some extra
resistance plateaus in long graphene junctions. This is explained by
the suggestion that only part of edge states participate in the
mixing mechanism.

The rest of the paper is organized as follows. In Section
\ref{sec2}, we describe the model and give the details of our
calculations. In Section \ref{sec3}, the numerical results are
given. Finally, a brief conclusion is presented in Section
\ref{sec4}.

\section{Model and calculation}

\label{sec2}

We consider the six-terminal graphene junction device which is
illustrated in Fig.~\ref{fig1}(a). The six terminals are labelled by
the terminal-$i$ ($i=1,2,...,6$). Terminals 1 and 4 are the current
source and drain, respectively, and other terminals are used as the
voltage probes. The whole graphene device is basically divided into
three regions: the left, central and right regions as shown in
Fig.~\ref{fig1}(a). Each of the left and right regions includes
three terminals. The dimension of the central region is described by
the integers $N$ and $M$ [see the red-site (dark gray) region in
Fig.~\ref{fig1}(a)]. There are totally $2N\times (2M+1)$ carbon
atoms in the central region. The width and length of the central
region are $a(3N-1)$ and $\sqrt{3}a(M+1)$, respectively. The width
of voltage terminals is chosen as $\sqrt{3}aN$, where $a\approx
0.142$ nm is the distance between two neighboring carbons.
Fig.~\ref{fig1}(a) shows the case $N=3$ and $M=10$.

The whole system is subjected to a perpendicular magnetic field
which leads to the formation of Landau levels. In quantum Hall
regime, bulk states are compressible, and the chiral edge states
flow along the edges. This behavior is independent of the type of
the edges if the graphene ribbon is wide enough. So we choose wide
zigzag-graphene ribbon for our simulation. Compared to the hopping
term, both Zeeman splitting and spin-orbit coupling are very small
and hence they are negligible. Furthermore, we adopt Anderson
on-site disorder. In fact, other types of disorders (e.g. the
interface disorder and the long-range disorder) may exist in the
experimental device and also lead to the edge state mixing. But from
the point of the edge state mixing, these other types of disorders
should have the similar effect. Here we can assume Anderson disorder
only exists in the central region based on the fact that the effect
of disorder in the left and right regions is suppressed by edge
states when the ribbon is wide enough.\cite{note1}

In the tight-binding representation, the Hamiltonian of the six-terminal
graphene junction is given by\cite{prl101-166806, prb78-205308, prb73-233406}
\begin{equation}
H=\sum_{i}{\epsilon _{i}a_{i}^{\dagger }a_{i}}-\sum_{<ij>}{(te^{i\phi
_{ij}}a_{i}^{\dagger }a_{j}+\emph{h.c.})},
\end{equation}
where $a_{i}^{\dagger }$ and $a_{i}$ are respectively the creation
and annihilation operators at site $i$, and $\epsilon _{i}$ the
energy of Dirac point (i.e., the on-site energy). In the left and
right regions, $\epsilon _{i}$ is equal to $E_{L}$ and $E_{R},$
respectively [Fig.~\ref{fig1}(b)], which can be controlled by the
gate voltage in the experiment. In the central region, $\epsilon
_{i}=k(E_{R}-E_{L})/(2M+2)+E_{L}+w_{i}$, where the column index
$k=1,2,...,2M+1$ [see Fig.~\ref{fig1}(a)] and $w_{i}$ is the on-site
disorder energy. $w_{i}$ is assumed to be randomly distributed in
the range $[-W/2,W/2],$ where $W$\ is the disorder strength. The
second term in the Hamiltonian stands for the nearest-neighbor
hopping. The effect of the magnetic field $B$ is addressed by the
phase $\phi _{ij}=\int_{i}^{j} \vec{A}\cdot d\vec{l}/\phi _{0}$ in
the hopping interaction Hamiltonian where $\vec{A}=(-By,0,0)$ is the
vector potential and $\phi _{0}=\hbar /e$. The magnetic field $B$ is
applied to the whole device (including the six terminals and central
region) along the perpendicular direction.

The multi-terminal resistance \cite{prb38-9375} is defined as $%
R_{ij,kl}=(V_{k}-V_{l})/I_{i\leftarrow j}$, where the contacts $i$ and $j$
are terminals used to draw and input current, and the two contacts $k$ and $%
l $ are used to measure the voltage difference. We introduce two
longitudinal resistances $R_{t}=R_{14,23}$ (on the top edge) and $%
R_{b}=R_{14,65}$ (on the bottom edge) and two Hall resistances $%
R_{l}=R_{14,26}$ (in the left region) and $R_{r}=R_{14,35}$ (in the right
region). These four resistances obey the relation,
\begin{equation}
R_{t}+R_{r}=R_{b}+R_{l}.  \label{eq2}
\end{equation}

From the Landauer-B\"{u}ttiker formula at zero temperature, the current
flowing into terminal-$i$ is given by $I_{i}=(2e^{2}/h)\sum_{j(\neq
i)}T_{ij}(E_{F})(V_{i}-V_{j})$.\cite{datta} Here, $T_{ij}(E_{F})=\text{Tr}[%
\mathbf{\Gamma }_{i}\mathbf{G^{r}\Gamma }_{j}\mathbf{G^{a}}]$ $%
(i,j=1,2,...,6 $ and $i\neq j)$ is the transmission coefficient from
terminal-$j$ to terminal-$i$ at Fermi energy $E_{F}$, $\mathbf{\Gamma }%
_{i}(E_{F})=i{[\mathbf{\Sigma }_{i}^{r}(E_{F})-\mathbf{\Sigma }%
_{i}^{a}(E_{F})]}$ the line width functions, $\mathbf{G}^{r}(E_{F})=[\mathbf{%
G}^{a}]^{\dagger }=1/[E_{F}-\mathbf{H}_{cc}-\sum_{i=1}^{6}\mathbf{\Sigma }%
_{i}^{r}]$ the retarded and advanced Greens functions, and
$\mathbf{H}_{cc}$ the Hamiltonian of the dashed-box region which
includes two green-site (light gray) regions and the red-site (dark
gray) central region [see Fig.~\ref{fig1}(a)]. The retarded
self-energy ${\mathbf{\Sigma }}_{i}^{r}(E_{F})$ due to the coupling
to the terminal-$i$ can be calculated numerically.\cite{jpf15-851}
To determine the longitudinal and Hall resistances mentioned above,
we applied a bias $V$ across terminal-1 and terminal-4, and the
currents in the voltage probes (terminals 2, 3, 5, and 6) are set to
zero. Then from the
Landauer-B\"{u}ttiker formula, the voltages $V_{2}$, $V_{3}$, $V_{5}$ and $%
V_{6}$ of the voltage probes and the currents $I_{1}$ and $I_{4}$ can be
calculated. We should have $I_{1}=-I_{4}\equiv I_{14}$. Finally, the
longitudinal and Hall resistances are given by $R_{t}\equiv
(V_{2}-V_{3})/I_{14}$, $R_{b}\equiv (V_{6}-V_{5})/I_{14}$, $R_{l}\equiv
(V_{2}-V_{6})/I_{14}$ and $R_{r}\equiv (V_{3}-V_{5})/I_{14}$.

The recursive Green's function technique \cite{jpf15-851} is used for the
computation of the transmission coefficient. The Fermi energy $E_{F}$ is set
equal to zero as the energy reference point. The hopping energy $t\approx $
2.75 eV is used as the energy unit, which corresponds to $3\times 10^{4}$ K.
It is reasonable to assume zero temperature condition in our calculations
because the temperature in the experiment is only of several Kelvin or
sub-Kelvin. Taking into account the spin degeneracy, we will use $h/2e^{2}$
as the resistance unit. The corresponding filling factors $\nu _{L}$ and $%
\nu _{R}$ are taken as odd integers ( $\pm 1,\pm 3,\pm 5,...)$ instead of
even integers ($\pm 2,\pm 6,\pm 10,...)$. The effect of the constant
magnetic field is addressed by appropriate Peierls phase \cite{prb77-115408}%
: $2\phi \equiv (3\sqrt{3}/2)a^{2}B/\phi _{0}$, where $a\approx
0.142$ nm is the distance between two neighboring carbons. We will
take $\phi =0.01$ which corresponds to the magnetic length
$l_{B}=\sqrt{\hbar /(eB)}\approx 1.6 $ nm.\cite{note2} We will
consider the case $N=50$ and $M=20$, where the area of the central
region is $21.2\times 5.2$ nm$^{2}$ and the width of the voltage
terminals is $12.3$ nm. The reason for using a width of the graphene
ribbon far larger than the magnetic length is that edge states can
not mix except near the boundary of the junction. Finally, the
disorder is averaged over 2000 random configurations except in
Fig.~\ref{fig4}(a) and (b), where 500 configurations are taken.

\begin{figure}[tbp]
\centering
\includegraphics[bb=72 46 897 696,width=8.5cm,totalheight=7cm]{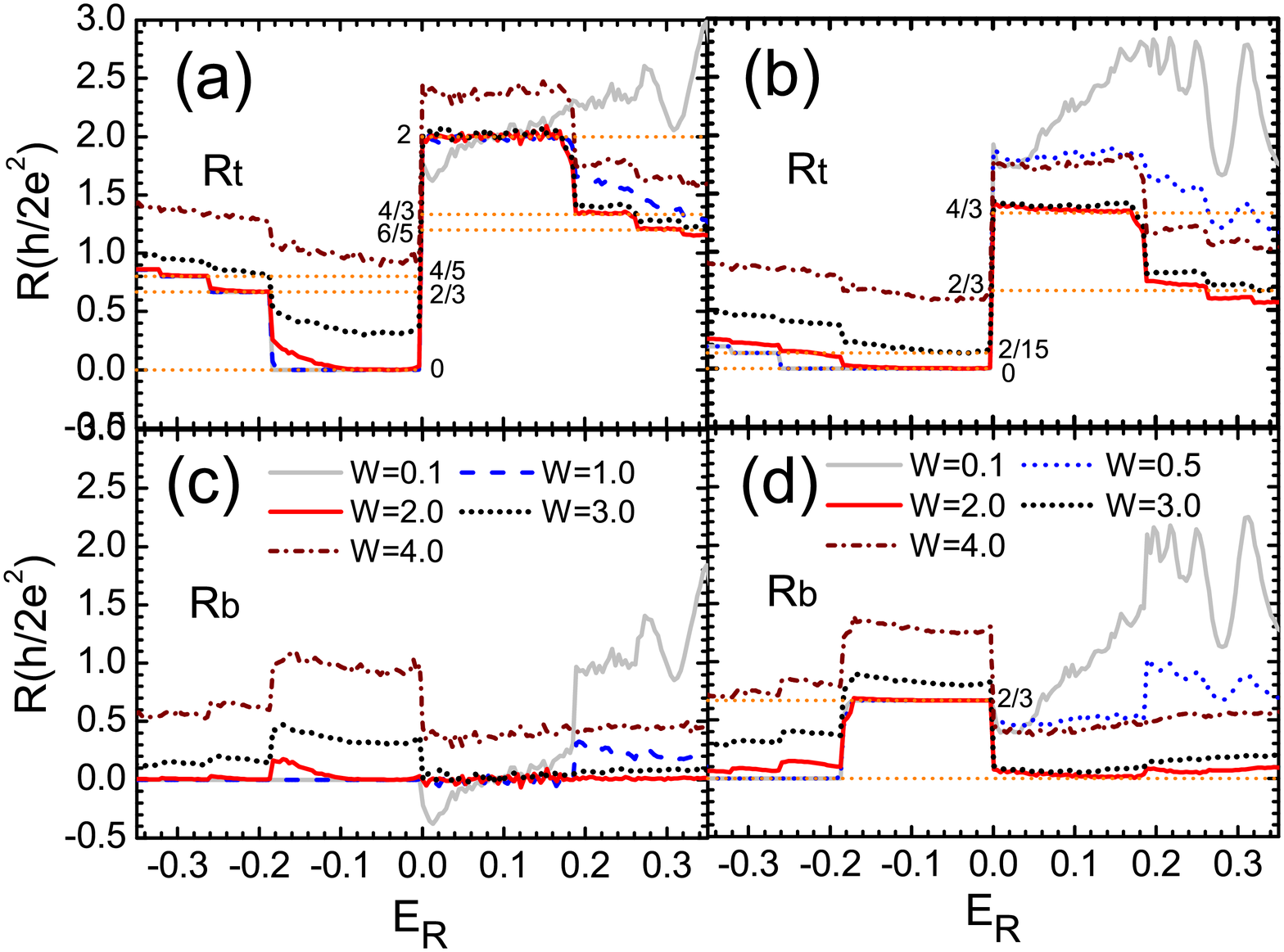}%
\newline
\caption{(Color online) The longitudinal resistances $R_t$ and $R_b$ vs. $%
E_R $ for the different disorder strength $W$ at the energy $E_L=-0.1$
[panels \textbf{(a)} and \textbf{(c)}] and $-0.2$ [panels \textbf{(b)} and
\textbf{(d)}]. The panels \textbf{(a)} and \textbf{(b)} are for $R_t$ and
the panels \textbf{(c)} and \textbf{(d)} are for $R_b$. The other parameters
are the same as in Fig.~\protect\ref{fig1}(c) and (d).}
\label{fig2}
\end{figure}

\section{Numerical results and analysis}

\label{sec3}

We first study the resistances in the clean graphene junction under
a strong magnetic field with $\phi =0.01$. In Fig.~\ref{fig1} (c)
and (d), the Hall and longitudinal resistances ($R_{l}$, $R_{r}$,
$R_{t}$, and $R_{b}$) versus the Dirac energy $E_{R}$ of the right
region are shown. The Dirac energy $ E_{L}$ of the left region is
fixed at $-0.1$ [Fig.~\ref{fig1}(c)] or $-0.2$ [Fig.~\ref{fig1}(d)],
which corresponds to $\nu _{L}=1$ and $3$, respectively. As usual,
the Hall resistances $R_{l}$ and $R_{r}$ display quantized plateaus
with plateau values $\pm 1,~\pm 1/3,~\pm 1/5,$ ... [in units of
$h/2e^{2}$], and Hall plateau of a region only depends on the
filling factor of the region, $R_{l}=1/\nu _{L}$ and $R_{r}=1/\nu
_{R}$. Furthermore, the Hall resistance plateaus are found to be
unaffected by the sizes of the central region and the presence of
the disorder in the central region, because that the Hall effect is
very robust. We will then focus our study on the longitudinal
resistances $R_{t}$ and $R_{b}$.

In the presence of the magnetic field, the longitudinal resistance $R_{t}$
of the top edge usually is not equal to $R_{b}$ of the bottom edge
regardless of the junction type and disorder strength $W$, because Hall edge
states have the chirality which breaks the symmetry of the top and bottom
edges. With $E_{R}<0$, both $R_{t}$ and $R_{b}$ of the clean n-n junction
device exhibit perfect plateau structures [Fig.~\ref{fig1}(c) and (d)]. By
considering the carrier transport along the edges, the plateau values can be
analytically obtained:
\begin{equation}
R_{t}=\left( \frac{1}{|\nu _{L}|}-\frac{1}{|\nu _{R}|}\right) \frac{h}{2e^{2}%
}\quad \mathrm{\ and}\quad R_{b}=0  \label{eq3}
\end{equation}%
for the n-n$^{+}$ regime ($0<\nu _{L}\leq \nu _{R}$) and
\begin{equation}
R_{t}=0\quad \mathrm{\ and}\quad R_{b}=\left( \frac{1}{|\nu _{R}|}-\frac{1}{%
|\nu _{L}|}\right) \frac{h}{2e^{2}}  \label{eq4}
\end{equation}%
for the n$^{+}$-n regime ($0<\nu _{R}\leq \nu _{L}$). The plateau values can
be understood from the following simple argument. It is well known that, for
the n-n regime the carriers are electron-like and they move clockwise. For
the case $0<\nu _{L}\leq \nu _{R}$ (the n-n$^{+}$ regime), all edge states
at the bottom edge are from the right reservoir (i.e. terminal-4), so the
voltages $V_{6},V_{5}$ and $V_{4}$ are all equal and this leads to $%
R_{b}=(V_{6}-V_{5})/I_{14}=0$. According to Eq.~(\ref{eq2}), $R_{t}$ is then
equal to $(1/|\nu _{L}|-1/|\nu _{R}|)h/2e^{2}$. The numerical results in
Fig.~\ref{fig1}(c) and (d) are consistent with the plateau values given in
Eqs.(3) and (4). For example, in Fig.~\ref{fig1}(c) where $\nu _{L}=1$, the
resistance $R_{b}$ is zero and $R_{t}$ can be equal to $0$, $(2/3)h/2e^{2}$,
$(4/5)h/2e^{2}$, ... for $\nu _{R}=1$, $3$, $5$, .... And in Fig.~\ref{fig1}%
(d) where $\nu _{L}=3$, $R_{b}$ is $0$ or $(2/3)h/2e^{2}$ and $R_{t}$ can be
$0$, $(2/15)h/2e^{2}$, $(4/21)h/2e^{2}$, and so on.

When $E_{R}>0$ the device becomes a n-p junction. There is no plateau
structure for $R_{t}$ and $R_{b}$, according to Fig.~\ref{fig1}(c) and (d).
Both $R_{t}$ and $R_{b}$ are very large and they are almost equal.
Furthermore, the smaller the filling factors are, the larger the
longitudinal resistances are. For some values of $E_{R}$, $R_{t}$ and $R_{b}$
can be over $1000h/2e^{2}$ (about $13M\Omega $). This is because for the
case of clean n-p junction, the edge states in the left and right regions
have different chiralities and they are well separated in space. Hence, edge
states mixture can not occur and this leads to very large longitudinal
resistances.

\begin{figure}[tbp]
\centering
\includegraphics[bb=180 268 415 504,width=8.7cm]{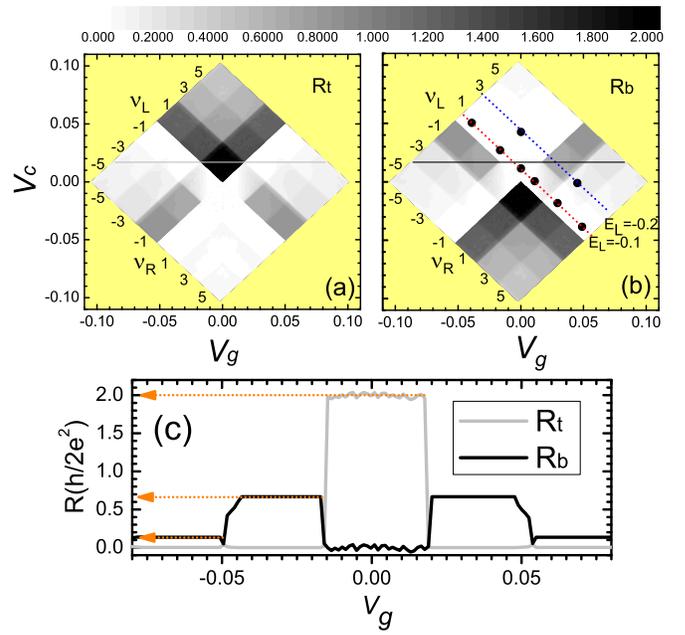}\newline
\caption{ (Color online) The 2-D plots of the longitudinal resistance $R_{t}$
(a) and $R_{b}$ (b) as a function of $V_{g}$ and $V_{c}$, with the
parameters $N=50$, $M=20$, $W=2$ and $\protect\phi=0.01$. (c) The
one-dimensional slices shown in (a) (gray solid) and (b) (black solid) are
plotted to show the dependence of $R_{t}$ and $R_{b}$ on $V_{g}$ with $V_{c}$
fixed at 0.015 and $W=1$. }
\label{fig3}
\end{figure}

Next, we shall discuss how $R_{t}$ and $R_{b}$ are affected by disorder.
Fig.~\ref{fig2} shows the dependence of $R_{t}$ and $R_{b}$ on the Dirac
energy $E_{R}$ for different disorder strength $W$. $E_{L}$ is fixed at $%
-0.1 $ ($\nu _{L}=1$) or $-0.2$ ($\nu _{L}=3$). For n-n regime ($E_{R}<0$),
the plateau structures of $R_{t}$ and $R_{b}$ still exist for weak and
moderate disorder strength. And the plateau values are the same as that of
the clean graphene junction. However, in the case of strong disorder (e.g. $%
W>3$), $R_{t}$ and $R_{b}$ increase and the plateau structures disappear.
This is expected as Hall edge states begin to be destroyed by strong
disorder. For n-p regime ($E_{R}>0$), $R_{t}$ and $R_{b}$ are strongly
reduced even when the disorder is weak. For example, when $W=0.1$ (very
weak) $R_{t}$ and $R_{b}$ are smaller than $3h/2e^{2}$ for any $E_{R}$ (Fig.~%
\ref{fig2}), though $R_{t}$ and $R_{b}$ can be over $1000h/2e^{2}$ at some
values of $E_{R}$ for clean junction [Fig.~\ref{fig1}(c)]. This significant
decrease results from that the electron-like and hole-like Hall edge states
start to mix in the vicinity of the n-p interface. It should be obvious from
our numerical result that the top-edge and bottom-edge resistances $R_{t}$
and $R_{b}$ are not equal and $R_{b}$ can be negative at some specific
values of the parameters [Fig.~\ref{fig2}(c)]. For suitable disorder
strength, the full edge-state mixture occurs and hence $R_{t}$ and $R_{b}$
exhibit plateau structures (see the curves of $W=2$ in Fig.~\ref{fig2}).
According to the Landauer-B\"{u}ttiker formula under the condition of full
edge-state mixture in the central region (junction), the plateau values can
be analytically obtained:
\begin{equation}
R_{t}=(\frac{1}{|\nu _{L}|}+\frac{1}{|\nu _{R}|})\frac{h}{2e^{2}}\mathrm{%
\quad and\quad }R_{b}=0.  \label{eq5}
\end{equation}%
In Fig.~\ref{fig2}, the plateau values for some low filling factors $%
(\nu_L,\nu_R)$ have been labeled and they are well consistent with the
numerical results.

\begin{figure}[tbp]
\centering
\includegraphics[bb=137 10 900 575,width=8.5cm]{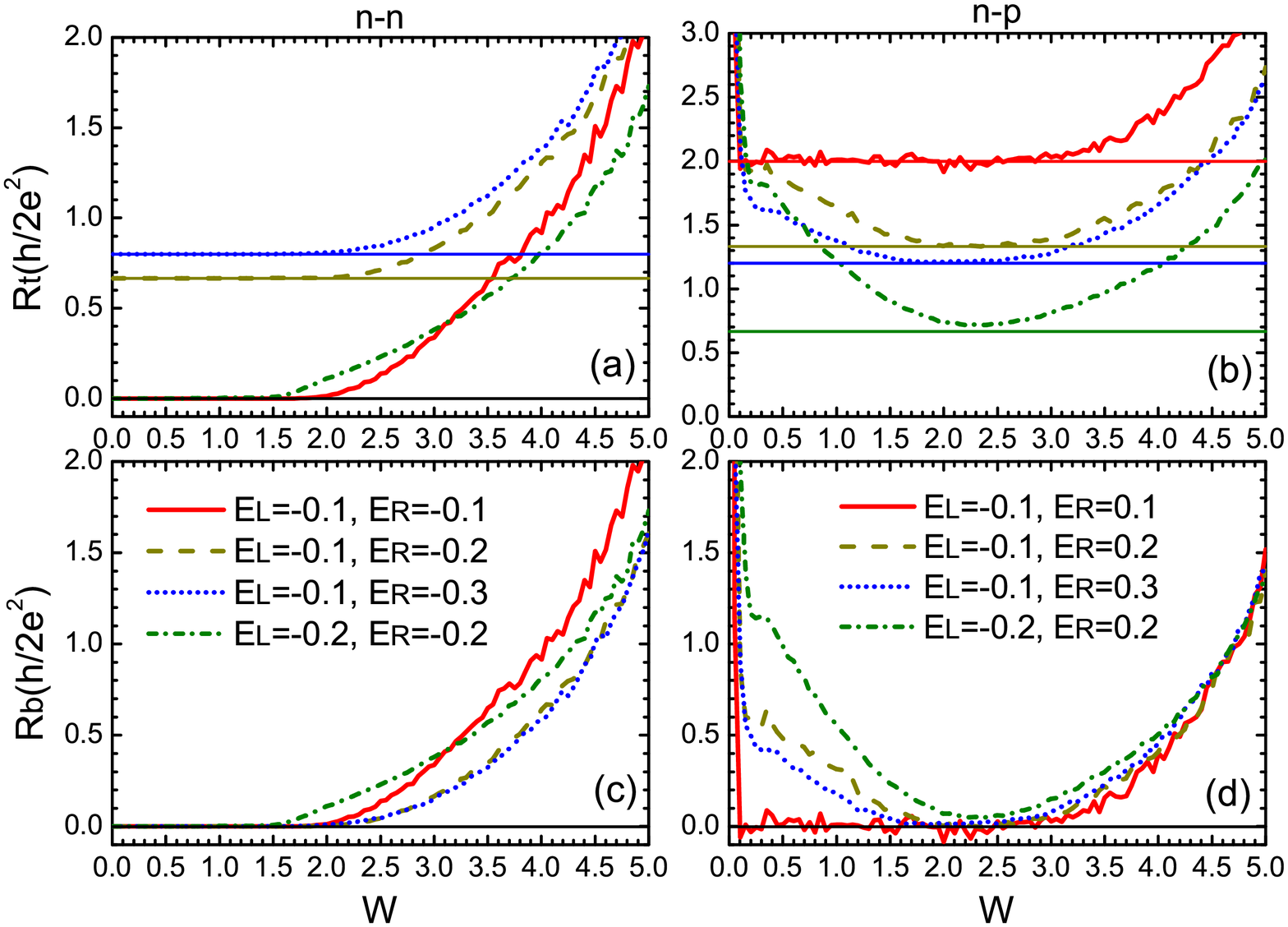}\newline
\caption{(Color online) The disorder dependence of longitudinal resistances $%
R_{t}$ and $R_{b}$ with selected $(\protect\nu _{L},\protect\nu _{R})$
[which are represented by the solid dots in Fig.~\protect\ref{fig3}(b)].
\textbf{(a)} and \textbf{(b)} are for $R_{t}$ and \textbf{(c)} and \textbf{%
(d)} are for $R_{b}$. The other parameters are the same as in Fig.~\protect
\ref{fig1}(c).}
\label{fig4}
\end{figure}

In the following we shall base on our model to simulate the recent
experimental results. In Ref.~\onlinecite{nanolett9-1973}, a difference of
carrier concentrations between the left and right regions was introduced by
chemical doping and a global gate voltage $V_{g}$ was used to control the
carrier concentrations of the whole region. So in our theoretical work, we
instead introduce two voltages $V_{g}$ and $V_{c}$ to control the carrier
concentrations. Let $n_{L/R}$ be the carrier concentration in the left/right
region. $n_{L}$ and $n_{R}$ are related to $V_{g}$ and $V_{c}$ in the
following way: $n_{L}\propto -V_{g}-V_{c}$ and $n_{R}\propto -V_{g}+V_{c}$.
Because of the linear dispersion relation of graphene, the carrier
concentration $n_{L/R}$ is approximatively proportional to $\mathrm{sgn}%
(E_{L/R})E_{L/R}^{2}$.\cite{nanaores1-361,note3} So we have
\begin{equation}
\mathrm{sgn}(E_{L})E_{L}^{2}=\beta (-V_{g}-V_{c}),\mathrm{sgn}%
(E_{R})E_{R}^{2}=\beta (-V_{g}+V_{c}),  \label{eq6}
\end{equation}%
where $\beta $ is a constant. We assign $\beta $ to be 1 in order to
simplify the expressions.

Figs.~\ref{fig3}(a) and (b) are the 2-D plots of $R_{t}$ and $R_{b}$ as
functions of $V_{g}$ and $V_{c}$ with the disorder strength $W=2$. The axes $%
E_{L}$ (or $\nu_L$) and $E_{R}$ (or $\nu_R$) shown in Figs.~\ref{fig3}(a)
and 3(b) are determined by Eq.(\ref{eq6}). The dotted lines in Fig.~\ref%
{fig3}(b) are identical to the curves plotted in Fig.~\ref{fig2}. From the
figure we see that $R_{t}$ and $R_{b}$ have the following symmetry
properties: (i) a mirror symmetry with respect to the line $V_{g}=0$
\begin{eqnarray}
R_{t}(\nu _{L},\nu _{R}) &=&R_{t}(-\nu _{R},-\nu _{L})  \label{eq7} \\
R_{b}(\nu _{L},\nu _{R}) &=&R_{b}(-\nu _{R},-\nu _{L}),
\end{eqnarray}
which reflects the inversion of the edge-state chiralities; (ii) an
inversion symmetry
\begin{equation}
R_{t}(\nu _{L},\nu _{R})=R_{b}(\nu _{R},\nu _{L})  \label{eq9}
\end{equation}
because of the interchange of $R_t$ and $R_b$ by rotating the angle $\pi$
round the center of the device. Furthermore, both resistances $R_t$ and $R_b$
exhibit the plateaus in the whole space of the parameters $V_g$ and $V_c$. $%
R_t$ and $R_b$ are approximatively constants at fixed filling factors $%
(\nu_L,\nu_R)$. But as $(\nu_L,\nu_R)$ varies, the jump of $R_t$ and $R_b$
occurs so that the borders of the filling factors $(\nu_L,\nu_R)$ are
clearly seen in Fig.~\ref{fig3}(a) and (b). Here the resistance plateau
values in n-n and n-p regimes are coincidental with Eqs.(\ref{eq3}), (\ref%
{eq4}) and (\ref{eq5}).

In Fig.~\ref{fig3}(c), the variations of $R_{t}$ and $R_{b}$ along the
horizontal solid lines in Fig.~\ref{fig3}(a) and (b) are shown. With the
range of the gate voltage $V_{g}$ from $-0.1$ to $0.1$, the corresponding
filling factors $(\nu _{L},\nu _{R})$ are $(-3,-5)$, $(-1,-3)$, $(1,-1)$, $%
(3,1)$, and $(5,3)$. The figure shows that in unipolar (n-n and p-p) regime,
the top-edge resistance $R_{t}$ is always equal to zero. But the bottom-edge
resistance $R_{b}$ is $(2/3)h/2e^{2}$ at $(\nu _{L},\nu _{R})=(-1,-3)$ and $%
(3,1)$, and $(2/15)h/2e^{2}$ at $(\nu _{L},\nu _{R})=(-3,-5)$ and $(5,3)$.
For the bipolar (n-p) regime, $(\nu _{L},\nu _{R})=(1,-1)$, the plateau of $%
R_{t}$ is $2h/2e^{2}$ and $R_{b}$ is zero. These resistance plateau values
are well in agreement with the recent experiment results [see Fig.5(c) in
Ref.~\onlinecite{nanolett9-1973}].

The disorder strength dependence of the longitudinal resistances at the
selected filling factors $(\nu _{L},\nu _{R})$ from Fig.~\ref{fig3}(b)
(represented by solid dots) is shown in Fig.~\ref{fig4}. Because of the
relation between $R_{t}$ and $R_{b}$ given in Eq.~(\ref{eq2}), $R_{t}$ and $%
R_{b}$ have similar characteristics with respect to the disorder strength $W$
even though $R_{t}\not=R_{b}$. For n-n regime (see the left panels of Fig.~%
\ref{fig4}), $R_{t}$ and $R_{b}$ in the clean junction ($W=0$) have plateau
structure. The plateau structure still exists when $W$ increases from zero.
However, when $W$ increases beyond a critical value $W_{c}$, $R_{t}$ and $%
R_{b}$ start to increase from the plateau values. The critical value $W_{c}$
is found to be dependent on the filling factors $(\nu _{L},\nu _{R})$: $%
W_{c}\approx 2.0$ for $(\nu _{L},\nu _{R})=(1,1)$, $(1,3)$ and $(1,5)$, and $%
W_{c}\approx 1.5$ for $(\nu _{L},\nu _{R})=(3,3)$. However, for n-p regime
(see the right panels of Fig.~\ref{fig4}), $R_{t}$ and $R_{b}$ are very
large in the clean junction. When $W$ increase from zero, $R_{t}$ and $R_{b}$
decrease sharply. For example, when $W$ increases from zero to 0.1, $R_{t}$
is reduced by two or three orders of magnitude to a finite value, while $%
R_{b}$ is reduced to zero. As $W$ continues to increase, both $R_{t}$ and $%
R_{b}$ decrease to certain plateau values. The plateau exists in a certain
range of disorder strength, where the full mixing of the electron-like and
hole-like edge state occurs. For the lowest filling factors $(\nu _{L},\nu
_{R})=(1,-1)$, the plateau values $R_{t}=h/e^{2}$ and $R_{b}=0$ exist in a
very broad range of disorder strength, from $0.1$ to $3.0$. For $(\nu
_{L},\nu _{R})=(1,-3)$ and $(1,-5)$, the disorder range for the existence of
resistance plateau are from 1.7 to 2.7 and from 1.9 to 2.7, respectively,
which is narrower than that of $(1,-1)$. For higher filling factors [$(\nu
_{L},\nu _{R})=(3,-3)$ or higher], the plateau does not exist, due to the
fact that it is more difficult to completely mix all edge states in the case
of high filling factors. This means that the resistance plateaus with lower
filling factors are easier to be observed in experiment. Finally, when $W$
further increase to be large than 3, $R_{t}$ and $R_{b}$ start to increase
for all cases of filling factors, which indicates that the edge states are
destroyed by strong disorder.

\begin{figure}[tbp]
\centering
\includegraphics[bb=68 11 953 475,width=8.5cm]{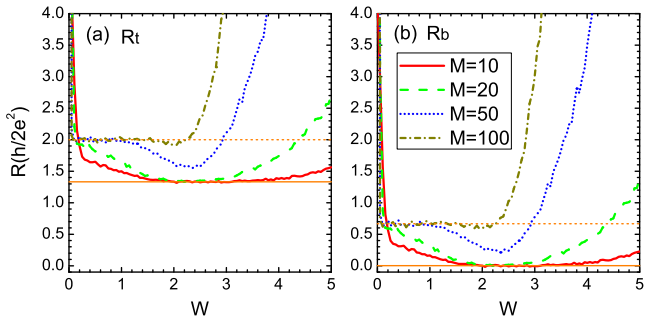}\newline
\caption{ (Color online) The resistances $R_t$ \textbf{(a)} and $R_b$
\textbf{(b)} vs. the disorder strength $W$ for the different length $M$ of
the center region. The parameters are $E_L=-0.1$, $E_R=0.2$, $\protect\phi%
=0.01$, and $N=50$.}
\label{fig5}
\end{figure}

Now we study the size effect of the central region on $R_{t}$ and $R_{b}$.
For unipolar graphene junction, the resistances $R_{t}$ and $R_{b}$ are
almost unaffected by the length $M$ and width $N$ of central region, except
for the case of very small $N$. In the following we focus on the bipolar
junction case. Fig.~\ref{fig5} shows the dependence of $R_{t}$ and $R_{b}$
on disorder strength $W$ at different length $M$ for the filling factors $%
(\nu _{L},\nu _{R})=(1,-3)$. For other filling factors, the results are
similar. From Fig.~\ref{fig5}, we can see that when the central region is
short (e.g. $M=10$), the full-mixing ideal plateaus [$R_{t}=(4/3)h/2e^{2}$
and $R_{b}=0$] exist in a wide disorder range $1.6<W<4$. This can be
understood as the small size of the central region makes the edge states
close to each other and hence the full mixing of the states occurs. With
increasing length $M$, the full mixing is more difficult. Under this
condition $R_{t}$ and $R_{b}$ are enhanced and the corresponding disorder
range for the ideal plateaus narrows down or even disappears. For example,
for $M=20$, the ideal plateau exists only in the disorder range $1.7<W<2.7$
which is much narrower than that of $M=10$. For larger $M$ such as $50$ and $%
100$, the ideal plateaus do not exist in any case of disorder strength.
Although full-mixing plateaus disappear in the case of longer junction,
extra plateaus [$R_{t}=2h/2e^{2}$ and $R_{b}=(2/3)h/2e^{2}$] emerge in a
long enough junction (e.g. $M=100$). The existence of the extra plateaus can
be explained by the partial mixing of edge states. The assumption of partial
mixing of edge states in long enough junctions is reasonable, because part
of the Hall edge states are near the junction boundary but the other Hall
edge states are far away from the boundary. Let us assume that there are $%
(x_{L},x_{R})$ edge states taking part in the full mixing [i.e., the
residual $(|\nu _{L}|-x_{L},|\nu _{R}|-x_{R})$ edge states in the left and
right regions do not participate in any mixing], the plateau values can be
analytically obtained by Landauer-B\"{u}ttiker formalism:
\begin{eqnarray}
R_{t} &=&(\frac{1}{x_{L}}+\frac{1}{x_{R}})\frac{h}{2e^{2}}  \nonumber
\label{eq10} \\
R_{b} &=&(\frac{1}{x_{L}}+\frac{1}{x_{R}}-\frac{1}{|\nu _{L}|}-\frac{1}{|\nu
_{R}|})\frac{h}{2e^{2}}
\end{eqnarray}
If $x_{L}=|\nu _{L}|$ and $x_{R}=|\nu _{R}|$ (i.e. full mixing), the result
of Eq.(\ref{eq10}) reduces to Eq.(\ref{eq5}). By taking $x_{L}=1$, $x_{R}=1$
and $(\nu _{L},\nu _{R})=(1,-3)$, the plateau values given by Eq.(\ref{eq10}%
) are $R_{t}=2h/2e^{2}$ and $R_{b}=(2/3)h/2e^{2}$, which are well consistent
with the numerical results in Fig.~\ref{fig5}.

Finally, we simply discuss how the width $N$ of a graphene junction affects $%
R_{t}$ and $R_{b}$. With increasing $N$ at fixed length $M$, the resistances
$R_{t}$ and $R_{b}$ decrease, the resistance plateaus exist in broader
disorder range regardless of the type of the junction (unipolar or bipolar).
Also, for the large $N$, some high-filling-factors resistance plateaus
emerge. This is expected as Hall edge states in a wider junction have more
chance to mix with each other.

\section{Conclusion}

\label{sec4}

In summary, we have investigated the longitudinal resistances at the two
edges of a six-terminal graphene junction in the presence of a perpendicular
magnetic field. By considering the presence of disorder in the vicinity of
the junction interface, the longitudinal resistances at opposite edges
exhibit different plateaus structures. It is found that the plateau values
are only determined by the Landau filling factors. The numerical results are
in excellent agreement with the recent experiment by Lohmann \emph{et al.}.
Furthermore, for unipolar junctions, resistance plateaus exist in clean
junctions. The plateau structure can be kept in the presence of weak and
moderate disorder, and they are destroyed by very strong disorder. However,
for bipolar junctions, the longitudinal resistances are very large and do
not have any plateau structure in the clean case. In the presence of
disorder, the resistances sharply drop even in the case of very weak
disorder and they exhibit plateau structures for suitable disorder strength.
In addition, we study the effect of the size of a graphene junction on the
resistances and find that some extra resistance plateaus emerge in long
graphene junctions. We explain this by proposing that only part of edge
states participate in the mixing.

\section*{Acknowledgements}

This work was financially supported by NSF-China under Grants Nos. 10734110,
10821403, and 10974236 and China-973 program.


\begin{references}
\bibitem{note1} If the disorder region is extended to the whole dashed-box
region, including the red-site (dark gray) central region and two
green-site (light gray) regions as shown in Fig.~\ref{fig1}(a), the
numerical results are almost unchanged. This is because the Hall
edge states are robust against the disorder.
\bibitem{note2}
$\phi=0.01$ corresponds $B\approx 250$T, which is quite a large
magnetic field. Due to the small size of the present system,  we
have to take a strong magnetic field to ensure the magnetic length
far smaller than the width of graphene ribbon. For a large system,
the magnetic field can be greatly decreased and the numerical
results remain the same.
\bibitem{note3}
Due to the lateral confinement of the system and the presence of the
magnetic field, the carrier concentration $n_{L/R}$ is not strictly
proportional to $\mathrm{sgn}(E_{L/R})E_{L/R}^{2}$. But it still is
a good approximation because the carrier concentrations $n_{L/R}$
only are the slight fluctuations around sgn$(E_{L/R}) E^2_{L/R}$. In
addition, this approximation does not change the value of the
resistance plateaus, except for slightly changing the width of the
resistance plateaus.
\end{references}
\end{document}